\documentclass[a4paper, 12pt]{article}
\usepackage{authblk}
\usepackage[bookmarksnumbered, plainpages]{hyperref}
\usepackage{hyperref}
\usepackage{amsmath, amsthm, amscd, amsfonts, amssymb, graphicx, color, booktabs}
\usepackage{subcaption}
\usepackage{braket}

\numberwithin{equation}{section}
\definecolor{email}{rgb}{0.00,0.00,0.84}

\begin{document}
\setcounter{page}{1}

\date{}

\title{\large \bf 12th Workshop on the CKM Unitarity Triangle\\ Santiago de Compostela, 18-22 September 2023 \\ \vspace{0.3cm}
\LARGE Summary of CKM 2023 Working Group~7: \lq\lq{}Mixing and CP violation in the $D$~system: $x_D$, $y_D$, $|q/p|_D$, $\phi_D$, DCPV in $D$ decays\rq\rq{}}

\author[1]{Patricia C.~Magalh$\tilde{\text{a}}$es \thanks{p.magalhaes@cern.ch}}
\author[2]{Tara Nanut Petri\v{c} \thanks{tara.nanut@cern.ch}}
\author[3]{Stefan Schacht \thanks{stefan.schacht@manchester.ac.uk}}
\affil[1]{Departamento de Raios Cósmicos e Cronologia, Universidade Estadual de Campinas 13083-860, Campinas, Brazil}
\affil[2]{European Organization for Nuclear Research (CERN), Geneva, Switzerland}
\affil[3]{Department of Physics and Astronomy, University of Manchester, Manchester M13 9PL, United Kingdom}
\maketitle

\begin{abstract}
We summarize the results of Working Group~7 at the \lq\lq{}12th International Workshop on the CKM Unitarity Triangle\rq\rq{} (CKM 2023) which took place in Santiago de Compostela, Spain, 18--22 September 2023.
\end{abstract} \maketitle

\section{Introduction}

Charm CP violation (CPV) is a unique probe of the flavor-structure of the up-quark sector.
This enables crucial probes for physics beyond the Standard Model (BSM) which are complementary 
to the ones from kaon and $b$ physics. A lot of progress both on the theoretical and the experimental side has been made and key results have been presented at CKM 2023.

On the experimental side, several experiments contribute to the results in the charm sector, each boasting their own strengths. In measurements of direct CPV, LHCb is dominating the high-precision studies due to the unmatched high statistics. The strength of Belle and Belle~II lies particularly in decays involving neutral and invisible particles, which are an essential piece in the puzzle for the theoretical interpretation of the data. For mixing and indirect CPV studies, LHCb again dominates the precision of measurements. BESIII crucially contributes to the effort by providing  dedicated measurements of strong phases as well as branching fraction measurements.
While accumulating new data and increasing the available statistical power is most crucial for future precision measurements, effort is constantly invested also into developing improved tools and techniques. Belle II presented their recently developed Charm Flavour Tagger tool~\cite{Bertemes:CKM23}, a novel method for the identification of the production flavour of neutral charmed mesons~\cite{Belle-II:2023vra}. It reconstructs particles that are most collinear with the signal meson and exploits the correlation of their charge with the flavour of the $D^0$. Compared to the previously predominantly used tagging method using $D^{*+}$ decays, the new technique doubles the signal statistics. A caveat is that more background is introduced, the level of which depends on the channel.

Due to the involved mass scales, the Standard Model (SM) prediction as well as the theoretical interpretation of charm decay data is challenging.
Recent key advances have been a systematic treatment of higher-order corrections in the $U$-spin expansion~\cite{Gavrilova:2022hbx}, 
employing experimental $\pi\pi$/$KK$ rescattering data \cite{Franco:2012ck, Bediaga:2022sxw, Pich:2023kim}
as well as a further development of LCSR techniques~\cite{Khodjamirian:2017zdu, Chala:2019fdb, Lenz:2023rlq}.
It is an open and on-going question if the large observed charm CP violation as well as the hints for large $U$-spin breaking can be explained with non-perturbative effects~\cite{Brod:2011re, Grossman:2019xcj, Schacht:2022kuj, Gavrilova:2023fzy} or are a sign of 
BSM physics, see for recent works Refs.~\cite{Dery:2019ysp, Bause:2020obd, Bause:2022jes}.

Below, we give an overview over the results of both theoretical and experimental advances in charm physics that were presented at CKM 2023 in Working Group 7.

\section{Direct CP Violation}

\subsection{Experiment }
Direct CP violation denotes CP violation in decay, and is measured in terms of the CP 
asymmetry
\begin{align}
a_{CP}^{dir}(D\to f) &\equiv \frac{
    \vert\mathcal{A}(D \to f)\vert^2 - \vert\mathcal{A}(\overline{D} \to \overline{f})\vert^2
    }{
    \vert\mathcal{A}(D \to f)\vert^2 + \vert\mathcal{A}(\overline{D} \to \overline{f})\vert^2
    }\,.
\end{align}
Direct CPV is a final-state-dependent phenomenon that results in an asymmetry of the squared matrix elements of CP-conjugated processes. For time-integrated CP asymmetries of $D^0$ decays, we have to a very good approximation~\cite{Grossman:2006jg}
\begin{align}
A_{CP} &\equiv \frac{
    \Gamma(D \to f) - \Gamma(\overline{D} \to \overline{f})}{
    \Gamma(D \to f) + \Gamma(\overline{D} \to \overline{f})} =  a_{CP}^{dir} + a_{CP}^{ind}\,,
\end{align}
\emph{i.e.},~the time-integrated CP asymmetry is the sum of direct and indirect CP asymmetries
$a_{CP}^{dir}$ and $a_{CP}^{ind}$, respectively. The indirect CP asymmetry is again a sum of a mixing induced asymmetry and one that stems from the interference between mixing and decay.

Experimentally, the variable that is being accessed is the raw asymmetry, which includes contributions both from the physical CP asymmetry $A_{CP}$ and from experimental nuisance asymmetries:
\begin{align}
A_{raw} = \frac{ N(D \to f) - N(\overline{D} \to \overline{f})}{
            N(D \to f) + N(\overline{D} \to \overline{f}))} = A_{CP} + \text{nuisance asymmetries}\,. 
\end{align}
The nuisance asymmetries depend on the experimental apparatus and the decay in question. They can include:
\begin{itemize}
    \item Production asymmetry: The asymmetry in the production of the charge-conjugated charm hadrons, or forward-backward asymmetries in production.
    \item Detection asymmetry: The asymmetry in the reconstruction efficiencies of positively and negatively charged particles. 
\end{itemize}
The general principles of dealing with nuisance asymmetries typically include subtracting or evaluating them through usage of normalisation channels, though the decay-particular methods are constantly evolving in light of the ever-increasing demand for precision.

\begin{figure} [t]
\centering
\includegraphics[width=9cm]{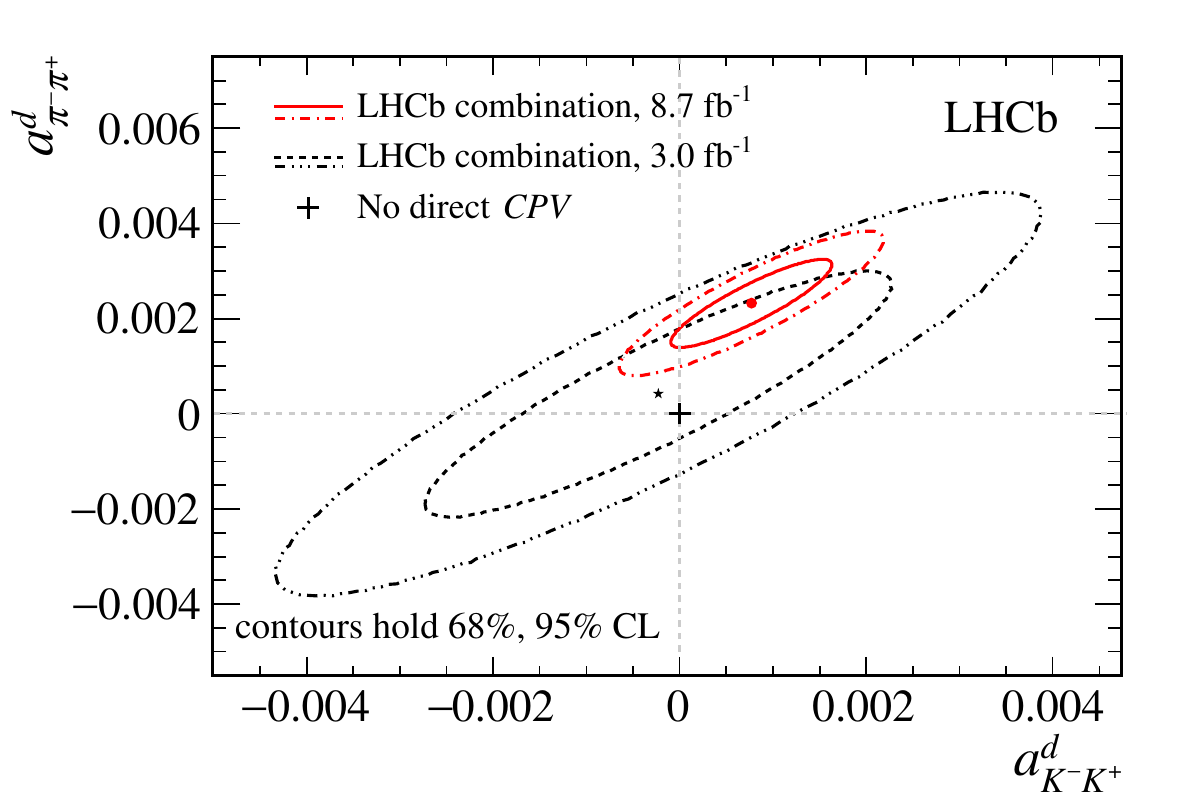}
\caption{Central values and two-dimensional confidence regions in the 
$\left(a^{dir}_{CP}(D^0\rightarrow K^+K^- ),a^{dir}_{CP}(D^0\rightarrow \pi^+\pi^-)\right)$ plane for the combinations of the LHCb results obtained with the dataset taken between 2010 and 2018 and the one taken between 2010 and 2012. Figure taken from Ref.~\cite{LHCb:2022lry}.}\label{ACP_KK_pipi}
\end{figure}

Undisputedly, the most high profile recent result regarding direct CP violation is the LHCb measurement of $A_{CP}$ in $D^0 \to K^+K^-$~\cite{LHCb:2022lry} using data collected in Run~2 (2015--2018, centre-of-mass energy $\sqrt{s} = 13$ TeV), and presented at CKM~2023 in Ref.~\cite{BrodzickaLHCb:CKM23}. Two calibration methods, each using a set of Cabibbo-favoured $D_{(s)}$ decays are employed to extract $A_{CP}$ from $A_{raw}$ at maximum statistical sensitivity. A combination of the two methods gives~\cite{LHCb:2022lry}
\begin{align}
    A_{CP}(D^0 \to K^+K^-) &= (6.8 \pm 5.4 \pm 1.6) \times 10^{-4}\,.
\end{align}
While this result is consistent with CP symmetry, when combining it with the measurement of $\Delta A_{CP} = A_{CP}(D^0 \to K^+K^-) - A_{CP}(D^0 \to \pi^+\pi^-) $  by LHCb~\cite{LHCb:2019hro}, the first observation of CP violation in the charm sector, it yields the first evidence of CP violation in the single channel of $D^0 \to \pi^+\pi^-$~\cite{LHCb:2022lry}:
\begin{equation}
    a^{dir}_{CP}(D^0\rightarrow \pi^+\pi^- ) = (23.2 \pm 6.1) \times 10^{-4}\,,
\end{equation}
as well as~\cite{LHCb:2022lry}
\begin{align}
    a^{dir}_{CP}(D^0\rightarrow K^+K^- ) &= (7.7 \pm 5.7) \times 10^{-4}\,.
\end{align}
A graphical representation of this result can be see in Fig.~\ref{ACP_KK_pipi}.

The interpretation of the result is subject to intense theoretical efforts, as it strongly breaks a U-spin symmetry relation, although yet only at the $2.7\sigma$ level~\cite{LHCb:2022lry}, see also Ref.~\cite{Schacht:2022kuj}. It remains unclear whether the result is compatible with the SM or is a sign of BSM physics, see the discussion in Sec.~\ref{directCPV-theory}. 

Further recent measurements of CP asymmetries include the LHCb study of decays $D^+\to\eta{(')} \pi^+$ and $D_s^+\to\eta{(')}\pi^+$ using Run~2 data~\cite{LHCb:2022pxf}, and the Belle study of $D^0\to K_S^0 K_S^0 \pi^+\pi^-$ using the full Belle dataset, collected on or near the
$\Upsilon(4S)$ and $\Upsilon(5S)$ resonances~\cite{Belle:2023qio}. 
All results are compatible with CP symmetry:
\begin{align}
    A_{CP}(D^+\to\eta \pi^+) & = (3.4 \pm 6.6 \pm 1.6 \pm 0.5) \times 10^{-3},  & &\cite{LHCb:2022pxf}\\
    A_{CP}(D^+\to\eta' \pi^+) & = (4.9 \pm 1.8 \pm 0.6 \pm 0.5) \times 10^{-3},& &\cite{LHCb:2022pxf}\\
    A_{CP}(D_s^+\to\eta \pi^+) & = (3.2 \pm 5.1 \pm 1.2 ) \times 10^{-3},& &\cite{LHCb:2022pxf}\\
    A_{CP}(D_s^+\to\eta'\pi^+) & = (0.1 \pm 1.2 \pm 0.8) \times 10^{-3},& &\cite{LHCb:2022pxf}\\
    A_{CP}(D^0\to K_S^0 K_S^0 \pi^+\pi^-) & = (-2.51 \pm 1.44^{+0.11}_{-0.10})\% \,.& &\cite{Belle:2023qio}
\end{align}
Besides measuring  $A_{CP}$, there exist further methods to search for direct CPV. Of special interest are those particular to multi-body decays that are sensitive to local effects, as the rich dynamics of multi-body decays with varying strong-phase differences across the decay phase space can increase the sensitivity to CPV in certain areas. Broadly, we can divide those into model-dependent and model-independent techniques. The model-dependent case represents amplitude analyses, which search for CP violation in each amplitude. Model-independent methods are often discovery techniques, which aim to check the consistency with CP symmetry but do not provide insights into the underlying CP-violating mechanisms. 

Several recent results using different model-independent techniques were presented at CKM~2023, and are detailed in the following paragraphs. 
$T$-odd triple product asymmetry measurements are complementary to  $A_{CP}$ measurements in regard to their dependence on the strong phase: while  $A_{CP} \propto \sin{\delta}$,  $a_{CP}^{T-odd} \propto \cos{\delta }$, where $\delta$ is the relevant strong phase difference. Belle has recently measured $T$-odd asymmetries for several decays:
\begin{align}
    a_{CP}^{T-odd}(D^0\to K_S^0 K_S^0 \pi^+\pi^-) & = (-1.95 \pm 1.42^{+0.14}_{-0.12})\%\,, & & \cite{Belle:2023qio}\\
    a_{CP}^{T-odd}(D^+ \to K^+ K_S^0 \pi^+\pi^-)  &= (\phantom{-}0.34 \pm 0.87 \pm 0.32)\%\,, & & \cite{Belle:2023bzn}\\
    a_{CP}^{T-odd}(D_s^+ \to K^+ K_S^0 \pi^+\pi^-)  &=(-0.46 \pm 0.63 \pm 0.38)\%\,, & & \cite{Belle:2023bzn}\\
    a_{CP}^{T-odd}(D^+ \to K^+ K^- K_S^0 \pi^+)  &=(-3.34 \pm 2.66 \pm 0.35)\%\,, &  & \cite{Belle:2023bzn}\\
    a_{CP}^{T-odd}(D^+ \to K^- K^+ \pi^+ \pi^0)  &=(\phantom{-}2.6 \pm 6.6 \pm 1.3 )\times 10^{-3}\,, & &\cite{Belle:2023str}\\
    a_{CP}^{T-odd}(D^+ \to K^+ \pi^- \pi^+ \pi^0)  &=(-1.3 \pm 4.2 \pm 0.1 )\times 10^{-2}\,, & &\cite{Belle:2023str}\\
    a_{CP}^{T-odd}(D^+ \to K^- \pi^+ \pi^+ \pi^0)  &=(\phantom{-}0.2 \pm 1.5 \pm 0.8 )\times 10^{-3}\,, & & \cite{Belle:2023str} \\
    a_{CP}^{T-odd}(D_s^+ \to K^+ \pi^- \pi^+ \pi^0)  &=(-1.1 \pm 2.2 \pm 0.1 )\times 10^{-2}\,, & & \cite{Belle:2023str}\\
    a_{CP}^{T-odd}(D_s^+ \to K^- K^+ \pi^+ \pi^0)  &=(\phantom{-}2.2 \pm 3.3 \pm 4.3 )\times 10^{-3}\,, & & \cite{Belle:2023str}
\end{align}
Belle also made the first observation of the Cabibbo-suppressed decay $D_s^+ \to K^+K^-K_S^0\pi^+$~\cite{Belle:2023bzn}. All $a_{CP}^{T-odd}$  results are consistent with CP symmetry. 
As pointed out in Ref.~\cite{Bertemes:CKM23}, for the sub-region of phase space of $D_s^+ \to K^+h^-\pi^+\pi^0$ around the pseudo two-body decay $D_s^+ \to K^{*0}\rho^+$, it is found a large $T$-odd asymmetry of $a_{CP}^{T-odd} = (6.2 \pm 3.0 \pm 0.4)\%$~\cite{Belle:2023str}.\\

Another discovery technique is the energy test~\cite{Williams:2011cd, Parkes:2016yie}, which is an unbinned method for the statistical comparison of two distributions, \emph{e.g.}~the $D$ and $\overline{D}$ phase space. The test statistics is based on the distance of event pairs $(ij)$~\cite{BrodzickaLHCb:CKM23}:
\begin{equation}
    T = \frac{1}{2n(n-1)} \sum_{i,j\neq i}^{n} \psi_{ij} + \frac{1}{2\overline{n}(\overline{n} -1)}  \sum_{i,j\neq i}^{\overline{n}} \psi_{ij} - \frac{1}{n\overline{n}} \sum_{i,j\neq i}^{n,\overline{n}} \psi_{ij},
\end{equation}
where the first term represents the average distance in the $D$ sample, the second term the average distance in the $\overline{D}$ sample, and the third term the one between pairs in a combined sample.  The measured $T_0$ value is compared with a $T$-distribution from a CP-symmetrical sample, obtained with large numbers of permutations. Finally, a $p$-value is computed as $n(T>T_0)/n$ for a no-CPV hypothesis~\cite{BrodzickaLHCb:CKM23}. LHCb has conducted a search for CP violation in $D^0\to \pi^+\pi^-\pi^0$ decays using the energy test on data collected in Run~2~\cite{LHCb:2023mwc}. Particular motivation to study this channel is given by the fact that the dominant amplitudes of the underlying pseudo two-body decays proceed via the same quark-level transitions as the decay $D^0\to\pi^+\pi^-$. A $p$-value of 62\% is found, which is consistent with CP symmetry. Furthermore, LHCb used the energy test method to search for CPV in $D^0 \to K_S^0 K^{-}\pi^{+}$ and $D^0\to K_S^0 K^+\pi^-$, resulting in $p$-values of 70\% and 66\%, respectively, consistent with CP symmetry.\\

A complementary binned model-independent technique is the Miranda method. It constructs local-CPV observables in each Dalitz bin $i$, $S_{CP}^i$, which represent the significance of the difference between the two charge-conjugated $D$ samples. With no CPV present, it follows the standard normal distribution. A $\chi^2$ test is performed to evaluate the hypothesis. LHCb has used the Miranda method to search for CPV in $D_{(s)}^+ \to K^+K^-K^+$ decays~\cite{LHCb:2023qne}. The resulting $p$-values of 32\% ($D^+$) and 13\% ($D_s^+$) are consistent with CP symmetry.\\

A novel method is proposed for CPV searches, using the Earth mover’s distance as a measure of the distance between events in two samples~\cite{Davis:2023lxq}. The method has been presented at CKM 2023 in Ref.~\cite{Youssef:CKM23}.
Ref.~\cite{Davis:2023lxq} shows that this method can reach a similar sensitivity to CP violation as the energy test.  An additional benefit is the interpretability of the results, as these include also phase-space information regarding the localization of CP violation within the Dalitz plot. \\

In the charm baryon sector, Belle performed the first measurement of direct CP asymmetries in two-body singly Cabibbo-suppressed decays of charmed baryons~\cite{Belle:2022uod}:
\begin{align}
    A_{CP}(\Lambda_c^+ \to \Lambda K^+)   &= 0.021 \pm 0.026 \pm 0.001\,, \\
    A_{CP} (\Lambda_c^+ \to \Sigma^0 K^+) &= 0.025 \pm 0.054\pm 0.004\,,
\end{align}
consistent with CP symmetry. LHCb performed an amplitude analysis of $\Lambda_{c}^+ \to p K^-\pi^+$, presented at CKM 2023 in Ref.~\cite{Ukleja:CKM23}. The results can be used as input in future model-independent CPV searches~\cite{LHCb:2022sck}. The  average $\Lambda_{c}^+$  polarisation was measured to be large, $\mathcal{O}(65\%)$. A further LHCb analysis of the $\Lambda_{c}^+$ polarimetry produced a mapping of the $\Lambda_{c}^+$ polarisation across the whole Dalitz space~\cite{LHCb:2023crj}. The results can be used to probe $b\to c$ transitions
in semileptonic $\Lambda_{b}$ decays.

\subsection{Theory \label{sec:theory-direct-cp}}
\label{directCPV-theory}

The theoretical interpretation of charm CP violation is challenging, as no clear expansion parameter is available like in kaon and $B$ physics. The reason is the intermediate mass of the charm quark relative to the scale of QCD and the related possible long-distance effects.
A very promising strategy is to use the approximate SU(3)$_F$ symmetry of QCD,~\emph{i.e.}, 
the approximate invariance of the QCD Lagrangian under unitary rotations of the up, down and strange quark. Predictions based on this methodology include in particular sum rules between 
various CP asymmetries, which can be used in order to test the SM.
As was pointed out at CKM 2023 in Ref.~\cite{Nierste:CKM23}, these sum rules can be used systematically 
in order to probe the nature of contributions from BSM physics, namely whether it has a $\Delta U=0$ or $\Delta U=1$ transformation behaviour under $U$-spin.
Importantly, the consistency of the methodology can also be checked with a sum rule that holds for both $\Delta U=0$ and $\Delta U=1$ BSM physics~\cite{Nierste:CKM23}.

In the SM, it is expected that the $U$-spin limit relation
\begin{align}
a_{CP}^{dir}(D^0\rightarrow K^+K^-) + a_{CP}^{dir}(D^0\rightarrow \pi^+\pi^-) &= 0 \label{eq:U-spin-limit-sum-rule}
\end{align}
is violated at about $30\%$ due to $U$-spin breaking corrections.
The experimental data deviates from Eq.~(\ref{eq:U-spin-limit-sum-rule}) at $\sim 2.7\sigma$~\cite{LHCb:2022lry}, see Fig.~\ref{ACP_KK_pipi}. 
If confirmed, it is expected that further SU(3)$_F$ sum rules are also violated~\cite{Schacht:2022kuj, Grossman:2012ry, Grossman:2013lya, Muller:2015rna}, possibly revealing patterns in global analyses~\cite{Muller:2015rna, Hiller:2012xm, Muller:2015lua}.

A BSM scenario assuming NP in $c\rightarrow u\bar{d}d$ transitions could result \emph{e.g.}~in \cite{Nierste:CKM23}
\begin{align}
a_{CP}^{dir}(D^0\rightarrow \pi^+\pi^-) &= 3 a_{CP}^{dir}(D^0\rightarrow K^+K^-)\,, 
\end{align}
which is in agreement with the data. Therefore, a future more precise check of Eq.~(\ref{eq:U-spin-limit-sum-rule}) is essential.

As was shown in Ref.~\cite{Bause:2022jes} and presented at CKM 2023~\cite{Gisbert:CKM23}, beyond the SM, the data shown in Fig.~\ref{ACP_KK_pipi} can be explained with a leptophobic $Z'$ below $\mathcal{O}(20\, \mathrm{GeV})$. The corresponding fermion charges under the additional $U(1)'$ depend on the generation. Such theories can generate additional operators with the flavor content $\bar{s}c\bar{u}s$ and $\bar{d}c\bar{u}d$ with non-universal coefficients, thereby giving rise to
BSM $\Delta U=1$ transitions.
Furthermore, the presence of a $Z'$ would entail a pattern of CP violation in $D\rightarrow \pi\pi$ 
decays~\cite{Gisbert:CKM23} including the violation of the isospin limit prediction~\cite{Buccella:1992sg, Grossman:2012eb}
\begin{align}
a_{CP}^{dir}(D^+\rightarrow \pi^+\pi^0) &= 0\,.
\end{align}

Complementary to the searches for CP violation in meson decays is the phenomenology of 
CP violation in baryon decays. Baryon decays entail many additional opportunities highlighted in 
Ref.~\cite{Yu:CKM23}.  
In particular, models of rescattering can be systematically 
tested by confronting their predictions with data~\cite{Yu:2017zst}, giving important guidance to experiment.
A complication is the proliferation of hadronic parameters and the corresponding induced complexity. However, due to their non-zero spin, an important opportunity compared to meson decays is the construction of additional observables that have a complementary dependence on strong phases, namely $T$-odd and $T$-even correlations~\cite{Yu:CKM23}, see for details Refs.~\cite{Wang:2022fih, Durieux:2015zwa}.
Also for baryons a crucial technique will be the application of flavor symmetries, resulting \emph{e.g.}~in the $U$-spin limit sum rules~\cite{Grossman:2018ptn}
\begin{align}
A_{CP}(\Lambda_c^+\rightarrow p K^- K^-) + 
A_{CP}(\Xi_c^+\rightarrow \Sigma^+ \pi^-\pi^+) &= 0\,,\\
A_{CP}(\Lambda_c^+\rightarrow \Sigma^+\pi^- K^+) + 
A_{CP}(\Xi_c^+\rightarrow p K^- \pi^+) &= 0\,,\\
A_{CP}(\Lambda_c^+\rightarrow p\pi^- \pi^+) + 
A_{CP}(\Xi_c^+\rightarrow \Sigma^+ K^-K^+) &= 0\,.
\end{align}
Notably, these sum rules relate channels with different initial states $\Lambda_c^+$ and $\Xi_c^+$, unlike Eq.~(\ref{eq:U-spin-limit-sum-rule}) which relates $D^0$ decays. The reason is that the $D^0$ is a $U$-spin singlet, whereas $\Lambda_c^+$ and $\Xi_c^+$ form a $U$-spin doublet.

\section{Indirect CP Violation, Lifetimes and Strong Phases}

\subsection{Experiment  \label{sec:indirect-exp}}

Indirect CP violation in charm decays is connected to the quantum mechanical effect of mixing between $D^0$ and $\bar{D}^0$. The corresponding mass eigenstates are defined as 
\begin{align}
\ket{D_{1,2}} &= p \ket{D^0} \pm \,q \ket{\bar{D}^0}\,.
\end{align}
The mass eigenstates have finite mass and width differences $\Delta m$ and $\Delta \Gamma$.
Mixing is therefore governed by two parameters:
\begin{align}
x&\equiv \frac{\Delta m}{\Gamma}\,, &
y&\equiv \frac{\Delta \Gamma}{\Gamma}\,,
\end{align}
where $\Gamma$ is the average decay width. 
The condition $|q/p| \neq 1$ is equivalent to CP violation in mixing.
Another possibility of indirect CPV is through the interference between a decay with and without mixing.

LHCb reported recently a new measurement of time-dependent CP violation in $D^0\to K^+K^-$ and $D^0\to \pi^+\pi^-$ decays which have contributions from both direct and indirect CPV~\cite{LHCb_CPV_charm}:
\begin{align}
A_{CP}(D^0 \to f, t) &= a_{CP}^{dir}(D^0\to f) + \Delta Y_f \frac{t}{\tau_{D^0}},
\end{align}
where $f= \pi^+\pi^-$ or $K^+K^-$ and $\tau$ is the $D^0$ lifetime.
The measurement has been presented at CKM 2023 in Ref.~\cite{Betti:CKM23}.
From the time-dependent slope, the parameter $\Delta Y_f$ can be determined.
The current final-state dependent LHCb combinations are~\cite{LHCb_CPV_charm}
\begin{align}
    \Delta Y_{K^+K^-}     &= (-0.3 \pm 1.3 \pm 0.3)  \times 10^{-4}\,,\\
    \Delta Y_{\pi^+\pi^-} &= (-3.6 \pm 2.4 \pm 0.4)  \times 10^{-4}\,.
\end{align}
$U$-spin symmetry implies an approximate universality relation~\cite{Kagan:2020vri}
\begin{align}
\Delta Y = \Delta Y_{K^+K^-} \overset{\text{U-spin}}{=} \Delta Y_{\pi^+\pi^-}\,. \label{eq:DeltaY-universality}
\end{align}
Assuming Eq.~(\ref{eq:DeltaY-universality}), it is found by LHCb~\cite{LHCb_CPV_charm}
\begin{align}
 \Delta Y  = (-1.0 \pm 1.1 \pm 0.3)\times 10^{-4}\,.
\end{align}
Although twice as precise as the previous world average, $\Delta Y$ is still compatible with zero within one standard deviation, see Ref.~\cite{Betti:CKM23} for an overview of the different experimental results.

LHCb also measured the  difference between the effective decay width $\hat{\Gamma}$ of the decays $D^0 \to f$, $f = K^+K^-, \pi^+\pi^-$, and $\Gamma$, the average decay width of the $D^0$~\cite{LHCbMixCharm}:
\begin{equation}
    y_{CP}^f = \frac{\hat{\Gamma} (D^0 \to f) + \hat{\Gamma}(\overline{D}^0 \to f)) }{2\Gamma} -1. \label{eq:yCPf}
\end{equation}
Replacing in Eq.~(\ref{eq:yCPf}) the denominator $2\Gamma$ by $\hat{\Gamma} (D^0 \to K^-\pi^+) + \hat{\Gamma} (\bar{D}^0 \to K^+\pi^-)$ introduces a correction $y_{CP}^{K\pi}$ that has to be subtracted in order to obtain $y_{CP}^f$~\cite{LHCbMixCharm}. Averaging over the two signal channels, the result is~\cite{LHCbMixCharm} 
\begin{equation}
    y_{CP}-y_{CP}^{K\pi} = (6.96 \pm 0.26 \pm 0.13) \times 10^{-3}\,, \label{eq:yCPresult}
\end{equation} 
which is four times more precise than the previous world average, see Ref.~\cite{Betti:CKM23} for details. 
The inequality $y_{CP}\neq y$ would indicate CP violation.
The result Eq.~(\ref{eq:yCPresult}) drastically improves the precision of the world average of $y$.

Coming to three-body decays, LHCb conducted a comprehensive analysis of $D^0\to K^0_{S} \pi^+\pi^-$ decays~\cite{Dkspp_LHCb}, employing a method meticulously optimized for assessing $\Delta m$. The measurement, which accommodates measurements of both mixing and CP violation parameters, was separately conducted on two statistically independent samples based on the production mode of the $D^0$ meson. The combined results are~\cite{Dkspp_LHCb}: 
\begin{align}
    x_{CP}   &= \left(3.97 \pm 0.46 \pm 0.29 \right) \times 10^{-3}\,,\\
    y_{CP}   &= \left(4.59 \pm 1.2 \pm 0.85 \right) \times 10^{-3}\,,\\
    \Delta x &= \left(-0.27 \pm 0.18 \pm 0.01 \right) \times 10^{-3}\,,\\
    \Delta y &= \left(0.20 \pm 0.36 \pm 0.13 \right) \times 10^{-3}\,.
\end{align}
The above parametrisation can be translated into results for the mixing parameters $x$ and $y$, as well as the CP violating parameters $\vert q/p\vert$ and $\phi$, see Ref.~\cite{Dkspp_LHCb} for the corresponding results, including the very first observation of a non-zero mass difference in the $D^0$ meson system, with a significance of seven standard deviations.

The results presented by LHCb on the topic of charm mixing and indirect CPV constitute a tremendous improvement in precision, and are limited by statistical uncertainties~\cite{Betti:CKM23}. New measurements on larger datasets in the future will thus further improve the precision of the related parameters, though the theoretical interpretation remains challenging at the present stage. \\ 

Despite data samples that are several orders of magnitude smaller than those at LHCb, as explained at CKM 2023 in Ref.~\cite{Gao:CKM23}, BESIII can uniquely explore the quantum coherence between $D^0$ and $\bar D^0$ produced at the $\Psi(3770)$ resonance in the electron-positron collider. Such measurements can constrain CP violation parameters or provide crucial input for high-precision analyses of mixing and indirect CPV in other experiments. This is the case for the measurements of strong phases, with new results recently published by BESIII. Of particular interest is the measurement of the strong phase difference $\delta^{K\pi}_D$~\cite{BESIII:2022qkh}. Using data from quantum correlated decays including a $D^0\to K^0_{S,L} \pi \pi$ tag, it is found~\cite{BESIII:2022qkh}
\begin{equation}
    \delta^{K\pi}_D = (187.6^{+8.9\,+\,5.4}_{-9.7\,-\,6.4})^{\circ}\,.
\end{equation}
The result provides input not only for studies in the charm sector, but eventually contributes also to the precision measurements of the CKM angle $\gamma$.\\

BESIII is also in a unique position to measure CP-even fractions in multi-body charm decays, which serve as inputs for measurements of $\gamma$ as well. Recent results for CP-even fractions are:
\begin{align}
    F_+ (D^0 \to \pi^+\pi^-\pi^+\pi^-) &= 0.735 \pm 0.015 \pm 0.005\,, & & \cite{BESIII:2022wqs} \\
    F_+ (D^0 \to K_{S}^0 \pi^+\pi^-\pi^0) &= 0.235 \pm 0.010 \pm 0.002\,, & &\cite{BESIII:2023xgh} \\
     F_+ (D^0 \to K^+K^-\pi^+\pi^-) &= 0.730 \pm 0.037 \pm 0.021\,. & & \cite{BESIII:2022ebh}
\end{align}

\begin{figure}
    \centering
    \begin{subfigure}{.45\textwidth}
        \centering
        \includegraphics[width=.9\linewidth]{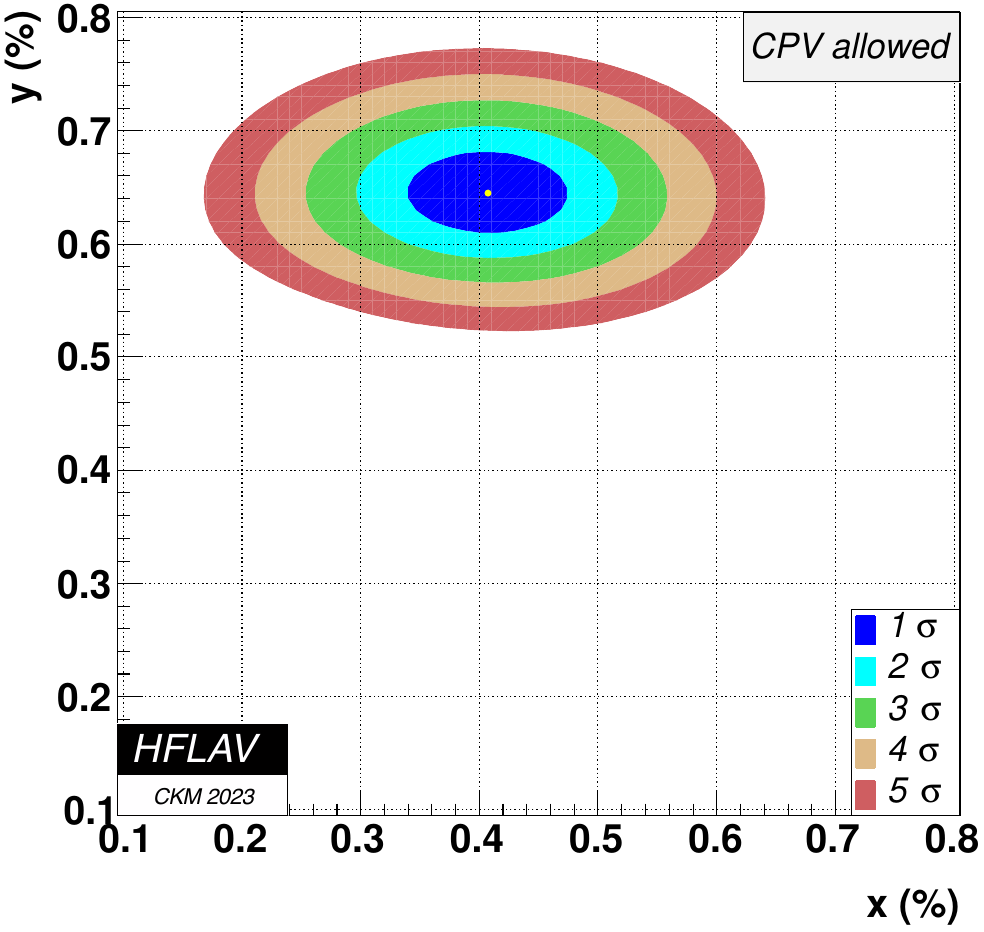}
        \caption{}
    \end{subfigure}%
    \begin{subfigure}{.45\textwidth}
        \centering
        \includegraphics[width=.9\linewidth]{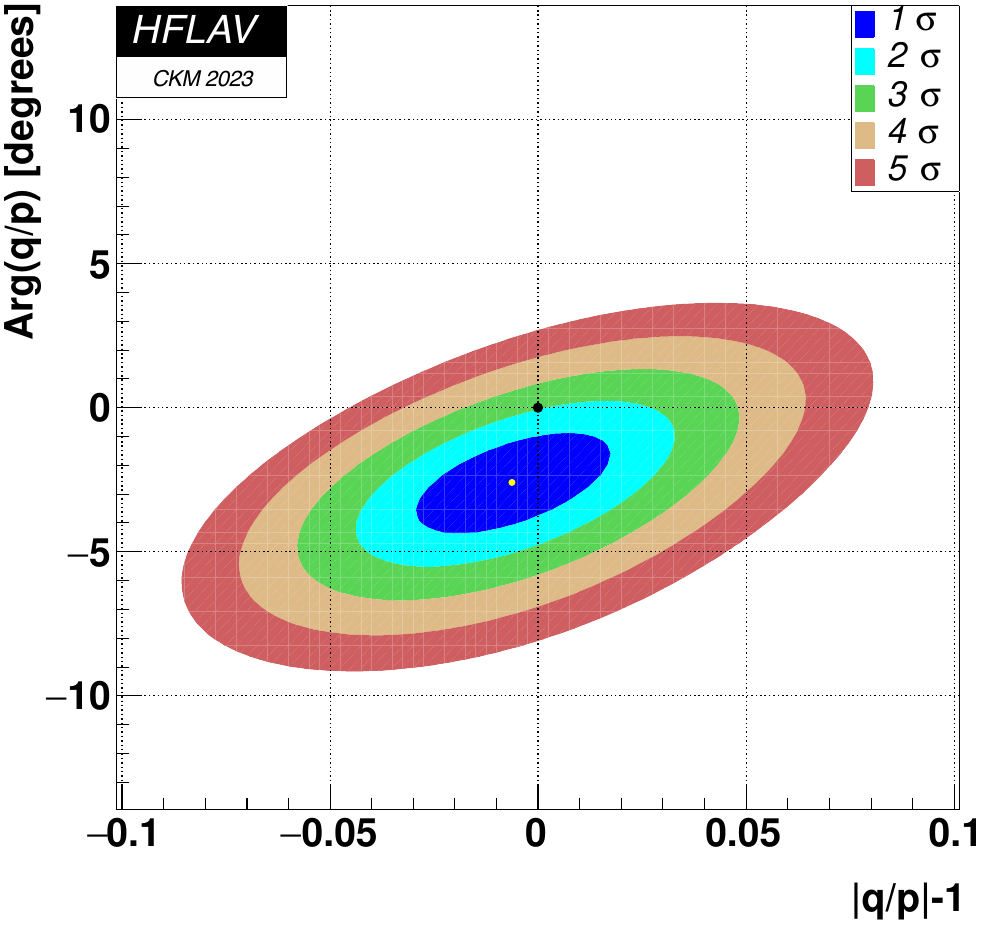}
        \caption{}
    \end{subfigure}
    \caption{HFLAV CKM 2023 update~\cite{HFLAV:2022esi} of the global fit for the mixing and CP violation parameters in charm decays. Figures taken from~Ref.~\cite{HFLAV:2022esi}.}
    \label{fig:HFLAVmixing}
\end{figure}

All the charm mixing and CP violation parameters for both direct and indirect CPV from different collaborations are gathered, processed and averaged by the Heavy Flavor Averaging Group (HFLAV). Their results have been updated with the discussed new measurements and presented at CKM~2023 in Ref.~\cite{BrodzickaHFLAV:CKM23}. The results for charm mixing are shown in Fig.~\ref{fig:HFLAVmixing} and read (fit with all CPV allowed)~\cite{HFLAV:2022esi}
\begin{eqnarray}
x &=& ( 0.407\pm 0.044)\%\,, \\
y&=& (0.645^{+0.024}_{-0.023})\%\,,\\
\phi&=& (-2.6^{+1.1}_{-1.2})^{\circ}\,,\\
|q/p| &=& 0.994^{+0.016}_{-0.015}\,.
\end{eqnarray}
Consequently, although mixing is well established in charm and $\phi$ is $2.2\sigma$ away from zero~\cite{BrodzickaHFLAV:CKM23}, we still cannot confirm indirect CPV.

Related to the topic, Belle II has produced world-leading measurements of the lifetimes of charm hadrons. The results are:
\begin{align}
    \tau (D^0) &= (410.5 \pm 1.1 \pm 0.8)~\mathrm{fs}\,,  & &\cite{Belle-II:2021cxx} \\
    \tau (D^+) &=  (1030.4 \pm 4.7 \pm 3.1)~\mathrm{fs}\,, & &\cite{Belle-II:2021cxx} \\
    \tau (\Lambda_c^+) &= (203.20 \pm 0.89 \pm 0.77)~\mathrm{fs}\,, & &\cite{Belle-II:2022ggx} \\
    \tau (D_s^+) &= (499.5\pm  1.7\pm 0.9)~\mathrm{fs}\,. & & \cite{Belle-II:2023eii}
\end{align}
Furthermore, Belle~II measured the lifetime of the $\Omega_c^0$ hadron,
\begin{align}
\tau(\Omega_c^0) &= (243\pm 48\pm 11)\, \mathrm{fs}\,,  \qquad \cite{Belle-II:2022plj}
\end{align}
which is consistent with results from LHCb~\cite{LHCb:2021vll}, and like these, differs from earlier measurements of experiments before the LHC.

\subsection{Theory \label{sec:indirect-th}}

In light of the big experimental advances regarding the measurements of mixing parameters and 
lifetimes, similar advances in the precision of theoretical predictions are important to 
interpret the current and future data.
These remain however challenging.

Regarding mixing, there are two main methodologies, the exclusive approach at hadron level, and the inclusive one based on the heavy quark expansion, see for details~\emph{e.g.}~the overview given in Ref.~\cite{Lenz:2020awd} and references therein.
In the exclusive approach one uses a complete set of possible hadronic intermediate states, relying on experimental input regarding the relevant branching ratios and strong phases.

\begin{figure}
    \centering
          \includegraphics[width=0.7\textwidth]{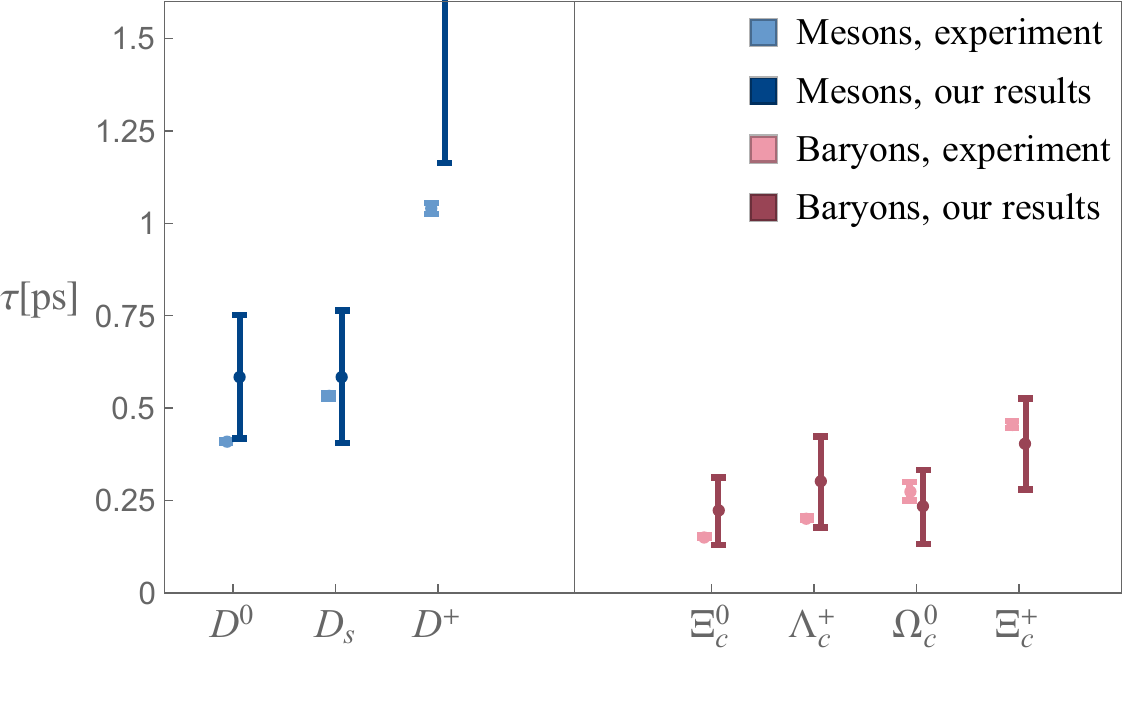}
        \caption{Comparison of theory predictions and experimental measurements of lifetimes. Figure taken from Ref.~\cite{Gratrex:2022xpm}. \label{fig:lifetimes}}
\end{figure}

In the inclusive approach, one employs the heavy quark expansion (HQE). Note that a necessary input here are the matrix elements from sum rules~\cite{Kirk:2017juj} or Lattice QCD~\cite{Carrasco:2014uya, Carrasco:2015pra, Bazavov:2017weg}.
At CKM 2023, in Ref.~\cite{Lenz:CKM23} it was analyzed in detail the possibility of the applicability of the 
heavy quark expansion to charm mixing and lifetimes.
It was stressed that the expansion in $\alpha_s$ and $\Lambda_{\mathrm{QCD}}/m_c$ is challenging, as~\cite{Lenz:CKM23}
\begin{align}
\alpha_s(m_c) &\sim 0.33\,, \qquad
\frac{\Lambda_{\mathrm{QCD}}}{m_c} \sim 3\, \frac{\Lambda_{\mathrm{QCD}}}{m_b}\,.
\end{align}
On top of that, big cancellations can take place due to the GIM mechanism, which is very strong in 
charm decays, compared for example to $b$ decays. The reason is that the relevant masses of the particles running in the loop $m_{d,s,b}$ are much closer to each other than $m_{u,c,t}$\,, which are the relevant ones for $b$ decays.
However, the amount of cancellations that occur might differ drastically for different charm observables. In fact, the lifetime of the $D^0$ can be predicted successfully in a $1/m_c$ expansion~\cite{Lenz:CKM23}. 
Due to large theoretical errors, heavy-quark predictions are also in agreement with the lifetime measurements of the charmed baryons, see Fig.~\ref{fig:lifetimes}. 
As a promising recent development, the same applies also to the mixing parameters~\cite{Lenz:2020efu}.
More work on higher order corrections of the heavy quark expansion is needed in order to improve the predictive power in the future.

\section{Conclusions}

Charm decays are a very active field right now, with a lot of current and on-going theoretical and experimental progress and great prospects in the near term. Experimental results have significantly improved the world average values of many variables. Since the previous CKM workshop, the most notable new results include the measurement of direct CPV in $D^0 \to K^+K^-$ decays by LHCb, which, when combined with previous measurements of $\Delta A_{CP}$, implies the first evidence of CP violation in a single decay channel, $D^0 \to \pi^+\pi^-$. Furthermore, the improvement of our experimental knowledge of charm mixing is astonishing: $(x,y)\neq (0,0)$ now beyond~$11.5\sigma$~\cite{HFLAV:2022esi}. Belle~II has published its first results in the charm sector with measurements of several charm hadron lifetimes with unprecedented precision. New results are expected in the near future, combining the Belle and Belle II datasets. BESIII has performed several measurements of strong phases, which provide crucial input for future analyses both in the charm sector and for the CKM angle $\gamma$.

On the theory side, further developments of flavor-symmetry methods are very promising for the goal to reach more precision for both meson and baryon decays, and new ideas have been explored in order to benefit from experimental rescattering data as well as from LCSR techniques. The heavy quark expansion has been successfully applied in order to predict charm lifetimes, and more work is needed in order to improve its predictions for charm mixing in the future. Recent tensions with SM expectations for direct CP violation have also led to a renewed interest in model-building. 

The further exploration of charm mixing and CP violation has a bright future and we are looking forward to the many great results that will be shown at the next CKM conference.\\

{\bf Acknowledgements.} 
We thank the speakers and the organizers of CKM~2023 for their excellent work. S.S.~is supported by a Stephen Hawking Fellowship from UKRI under reference EP/T01623X/1 and the STFC research grants ST/T001038/1 and ST/X00077X/1.

%\bibliographystyle{h-physrev}
%\bibliography{draft.bib}

\end{document}